\documentclass[lettersize,journal]{IEEEtran}
\usepackage{amsmath,amsfonts}
\usepackage{algorithmic}
\usepackage{algorithm}
\usepackage{array}
\usepackage[caption=false,font=normalsize,labelfont=sf,textfont=sf]{subfig}
\usepackage{textcomp}
\usepackage{stfloats}
\usepackage{url}
\usepackage{verbatim}
\usepackage{graphicx}
\usepackage{cite}
\def\K{\mathbb{K}}
\def\N{\mathbb{N}}

\def\th{\boldsymbol{\theta}}

\newtheorem{theorem}{Theorem}[section]

\newtheorem{remark}[theorem]{Remark}

\begin{document}

\title{Rescale-Invariant Federated Reinforcement Learning for Resource Allocation in V2X Networks}

\author{Kaidi Xu, Shenglong Zhou, and Geoffrey Ye Li,~\IEEEmembership{Fellow,~IEEE,}
\thanks{Kaidi Xu and Geoffrey Ye Li are with the ITP Lab, Department of EEE, Imperial College London, UK. Shenglong Zhou is with the School of Mathematics and Statistics, Beijing Jiaotong University, China. Emails:  k.xu21@imperial.ac.uk, shlzhou@bjtu.edu.cn, geoffrey.li@imperial.ac.uk.}
\thanks{*Corresponding author: Shenglong Zhou. This work was supported by the Fundamental Research Funds for the Central Universities and the Talent Fund of Beijing Jiaotong University.}
}



\maketitle

\begin{abstract}
Federated Reinforcement Learning (FRL) offers a promising solution to various practical challenges in resource allocation for vehicle-to-everything (V2X) networks. However, the data discrepancy among individual agents can significantly degrade the performance of FRL-based algorithms. To address this limitation, we exploit the node-wise invariance property of ReLU-activated neural networks, with the aim of reducing data discrepancy to improve learning performance. Based on this property, we introduce a backward rescale-invariant operation to develop a rescale-invariant FRL algorithm. Simulation results demonstrate that the proposed algorithm notably enhances both convergence speed and convergent performance.
\end{abstract}


\section{Introduction}
\IEEEPARstart{D}{eep}  reinforcement learning (DRL) is capable of optimizing long-term objectives for sequential decision-making problems without the requirement of a differentiable reward function and thus has been widely used for resource allocation in many  V2X communication systems \cite{ye2019deep,liang2019spectrum,he2020resource,
		nguyen2019distributed}.
		For example, a single agent model is learned in \cite{ye2019deep} based on the data from all vehicles and this agent sequentially makes decisions on power and sub-channel allocation for each vehicle-to-vehicle (V2V) link while when each V2V link is treated as an agent, a multi-agent DRL system has been formulated in \cite{liang2019spectrum}.
		
		However, when encountering multi-agents in V2X systems,  DRL-based algorithms often suffer from local observability of the vehicles and varying environment statistics caused by vehicular mobility. Performance degradation occurs when agents struggle to learn the dynamic environment based on the limited environment observations.
		To address these issues, federated reinforcement learning (FRL), the integration of federated learning (FL) and RL, provides a promising solution. It empowers each agent to learn the knowledge beyond its observability without raw data exchange, thus preserving privacy.  \cite{qi2021federated, lu2021dynamic, li2022federated}.
		For instance, FRL has been employed in \cite{lu2021dynamic} for distributed resource allocation in cellular networks to preserve privacy and alleviate communication overhead.  In \cite{li2022federated}, an FRL-based algorithm has been developed to optimize the cellular sum rate, reliability, and delay requirements of V2V links. The use of the FL framework is beneficial in overcoming limited observability and accelerating the training process.
	
	Despite the great potential of FRL, many FL-based algorithms, such as FedAvg \cite{mcmahan2017communication}, FedProx \cite{li2020federated}, and  FedGiA \cite{zhou2023fedgia},  face performance degradation when local devices hold significantly different data, including non-independent and identically distributed data or inhomogeneous data. Moreover, these algorithms pay much attention to the optimization model of a learning problem and thus do not make full use of the structures of neural networks. For example, ReLU-activated neural networks exhibit the node-wise invariance property \cite{neyshabur2015path}. By harnessing this feature, we can rescale the models for local devices to mitigate the influence of data differences and consequently enhance learning performance.
	
The major contribution of this paper lies in the development of the rescale-invariant FRL (RIFRL) algorithm for resource allocation in V2X networks. The algorithm fully exploits the node-wise invariance property of ReLU-activated neural networks, thereby remarkably improving the learning process of the FRL system under various configurations. Furthermore, the proposed backward rescale-invariant operation (BRIO) could prove advantageous in enhancing the performance of many distributed algorithms when addressing challenges posed by data discrepancy.

	\begin{table}\caption{Abbreviations. \label{tab:abbr}}
	\begin{tabular}{ll}\hline
	BRIO&Backward rescale-invariant operation\\ 
	DRL & Deep reinforcement learning\\
	FL & Federated learning\\
	FRL & Federated reinforcement learning\\
	PG & Policy gradient\\
	POMDP & Partially observable Markov decision process\\ 
	ReLU & Rectified linear unit \\
	RIFRL & Rescaling invariant FRL\\
	SINR & Signal-to-interference-plus-noise-ratio\\
	V2X & Vehicle-to-everything\\
	V2V & Vehicle-to-vehicle\\
	V2I & Vehicle-to-infrastructure\\  
	\hline
	\end{tabular}
\end{table}		
		
		\section{System Model and Problem Formulation}
To facilitate subsequent discussion, we also present all abbreviations in the sequel in Table \ref{tab:abbr}. Now the resource allocation for an orthogonal frequency-division multiple access based single-antenna V2X network is introduced as follows. 

In the V2X system, a roadside unit provides V2I link services to $N$ vehicles in the system and there are $K$ V2V links connecting neighbour vehicles. 
		Assume that each V2I link is allocated an orthogonal sub-channel, as it provides high-data-rate services to each vehicle. Therefore, the number of orthogonal sub-channels is the same as the number of the V2I links.
		The V2V links are enabled by device-to-device communication and reuse the uplink sub-channels of the V2I links to improve the spectral efficiency.
		By denoting the sets of V2I and V2V links as $\mathbb{N}:=\{1,2,\ldots,N\}$ and $\mathbb{K}:=\{1,2,\ldots,K\}$, respectively, V2I link $n$ is allocated with sub-channel $n$.
		
		At time slot $t$, V2I link $n$ receives the interference signals from V2V links. Therefore, the SINR for V2I link $n$ can be expressed as follows,
		\begin{equation}
			\gamma_n^{i,t}=\frac{|h_n^t[n]|^2P_n^{i,t}}{\sigma^2+\sum_{k\in\mathbb{K}}\rho_k^t[n]|\tilde{h}_k^t[n]|^2P_k^{v,t}},
		\end{equation}
		where $h_k^t[n]$ denotes the channel gain of V2I link $n$ and $\tilde{h}_k^t[n]$ represents the interference channel gain between the transmitter of V2V link $k$ and the road side unit in sub-channel $n$ at time slot $t$. $P_n^{i,t}$ and $P_k^{v,t}$ are the transmit power of V2I link $n$ and that of V2V link $k$ at time slot $t$, respectively. $\sigma^2$ is the variance of the additive Gaussian white noise. The binary variable, $\rho_k[n]^t$, represent whether sub-channel $n$ is allocated to V2V link $k$ for transmission, i.e., $\rho_k[n]^t=1$  if allocated, $\rho_k[n]^t=0$ otherwise. Thus the rate for V2I link $n$ at time slot $t$ can be expressed as,
		\begin{equation}
			R_n^{i,t} = W\log (1+\gamma_n^{i,t}),
		\end{equation}
		where $W$ denotes the bandwidth of each sub-channel.
		
		Similarly, at time slot $t$, V2V link $k$ receives the interference signals from other V2V and V2I links.
		The corresponding SINR for V2V link $k$ in sub-channel $n$ is given by,
		\begin{equation}
			\gamma_k^{v,t}[n]=\frac{\rho_k^t[n]|g_{k,k}^{v,t}[n]|^2P_k^{v,t}}{\sigma^2+I_k^{i,t}[n]+I_k^{v,t}[n]},
		\end{equation}
		where $g_{k,k}^{v,t}$ denotes the channel gain of V2V link $k$ at time slot $t$, $I_k^{i,t}[n]$ and $I_k^{v,t}[n]$ represent the received interference power at the receiver of V2V link $k$ from all V2I links and from other V2V links, respectively.
		Specifically, $I_k^{i,t}[n] = |\tilde{h}_{k,n}^t[n]|^2P_n^{i,t}$ and $I_k^{v,t}[n] = \sum_{k'\in \mathbb{K},k'\neq k}\rho_{k'}[n]|\tilde{g}_{k,k'}^{v,t}[n]|^2P_{k'}^{v,t}$, where $\tilde{h}_n^t[n]$ is the interference channel gain between the transmitter of V2I link $n$ and the receiver of V2V link $k$, and $\tilde{g}_{k,k'}^{v,t}[n]$ is that between the transmitter of V2V link $k'$ and the receiver of V2V link $k$ in sub-channel $n$.
		The rate  of V2V link $k$ is thus given by,
		\begin{equation}\begin{array}{l}
			R_k^{v,t}=\sum_{n\in\mathbb{N}}W\log(1+\gamma_k^{v,t}[n]).
		\end{array}\end{equation}
		
		V2V links are responsible for transmitting the periodically generated safety-related information. Therefore, we aim to allocate the transmit power and sub-channels to the V2V links to maximize the V2V package delivery rate within a given time duration for the V2X system without causing severe interference to the V2I links, i.e., $Pr(T_s\sum_{t\in\mathbb{T}} R_k^{v,t}>D_s),\forall k$, where $D_s$ denotes the size of data package, $T_s$ denotes the time duration of each time slot, and $\mathbb{T}:=\{1,2,\ldots, T\}$ denotes the set of transmission time indices. Note that the V2V link delivery rate is a long-term value, which can be effectively dealt with by RL as it suits sequential decision-making for a long-term goal. In addition, in the system, we also employ FL to combat local observability.

		\section{Framework of RIFRL}
		In this section,  we first formulate the resource allocation problem in the V2X system as a multi-agent POMDP  and then design the  RIFRL algorithm to solve the problem.
		\subsection{Multi-agent POMDP and PG for the V2X system}
		The designed V2X system can be formulated as a POMDP with a tuple {$\langle K, \mathbf{s}_t, \mathbf{a}_t^{(k)}, R_{t}^{(k)}, \mathbf{z}_{t}^{(k)}, P, O\rangle$}, where $K$ is the number of agents, $\mathbf{s}_t$ is the environment state at time  $t$, $\mathbf{a}_t^{(k)}$ is the action at time  $t$ of agent $k$, 
		{$\mathbf{z}_t^{(k)} = O(\mathbf{s}_t, k)$} is the local observation obtained by agent $k$, observation function $O(\cdot,\cdot)$ maps state $\mathbf{s}_t$ to specific observation $\mathbf{z}_t^{(k)}$ of agent $k$, $R_t^{(k)}$ is the local reward received by agent $k$ from the environment, and $P(\mathbf{s}_{t+1}\mid\mathbf{s}_t, \mathbf{a}_t)$ is a transition probability from state $\mathbf{s}_{t}$ with joint action {$\mathbf{a}_t=(\mathbf{a}_t^{(1)},\mathbf{a}_t^{(2)},\ldots,\mathbf{a}_t^{(K)})$} to next state $\mathbf{s}_{t+1}$. 
		In practice, we only use available {$\langle K, \mathbf{a}_t^{(k)}, R_{t}^{(k)}, \mathbf{z}_{t}^{(k)}\rangle$} to train the  system. Other quantities are automatically and implicitly learned by the agents during training.
				
		In the V2X system, each V2V link $k$ is deemed as an agent $k$ and makes decisions for itself.
		Therefore, we define 
		\begin{equation}
			\mathbf{a}_t^{(k)}:= \{P_k^{v,t},f_t^k\},
		\end{equation}
		where ${P_k^{v,t}\in\{23, 15, 5, -100\}}$ dBm is limited to discrete values for the sake of both ease of learning and practical circuit restriction and $f_t^k\in\mathbb{N}$ denotes the selected sub-channel index for agent $k$. (We restrict the number of selected sub-channels for each V2V link to 1.)
		We use the available information \cite{liang2019spectrum} as the observation information for each agent, i.e., 
		\begin{equation}
			\begin{aligned} 
				\mathbf{z}_t^{(k)}=\Big\{h_{n}^t[n],\tilde{h}_{k,n}^t[n],g_{k,k}^{v,t}[n],\tilde{g}_{k,k'}^{v,t}[n],I_{k}^{v,t-1}[n]+I_{k}^{i,t-1}[n],\\
				 \forall n\in{\N}, k'\in \mathbb{K},
				\mathbf{q}_k,T_k^t, B_k^t, k  \Big\},
			\end{aligned}
		\end{equation}
		where $\mathbf{q}_k$ is the relative position of the transmitter to the receiver of link $k$, $T_k^t$ and $B_k^t$ are the remaining time budget and the remaining data package size.
		Moreover, all agents share a common reward so that they can learn to cooperate.
		We give all the agents a reward of 0.5, once a V2V link successfully delivers its data package within the given time period.
		Therefore, considering the V2I link transmission quality, the reward for agent $k$ is given by,
		\begin{equation}\begin{array}{l}
			R_t^{(k)}=0.5*Q^t+0.1*\sum_n R_n^{i,t},
		\end{array}\end{equation}
	where $Q^t$ denotes the number of V2V links that complete data package transmission in time slot $t$.
		
		The general idea of POMDP is given as follows. At time $t$, each agent $k$ simultaneously selects and implements an action $\mathbf{a}_t^{(k)}$ based on its observation of the environment, $\mathbf{z}_t^{(k)}$. Subsequently, the environment returns a local reward, $R_t^{(k)}$, to each agent $k$ and it state $\mathbf{s}_{t}$ transits to the next state, $\mathbf{s}_{t+1}$, with a transition probability $P(\mathbf{s}_{t+1}\mid\mathbf{s}_t, \mathbf{a}_t)$. In the next time step, $t+1$, each agent $k$ obtains a new observation of the environment, $\mathbf{z}_{t+1}^{(k)}= O(\mathbf{s}_{t+1}, k)$. Throughout the paper, we set a common shared reward for all agents so that the targeted system is a Markov potential game.  According to \cite{leonardos2022global}, the PG-based algorithm for the game can finally converge to an approximate Nash equilibrium.
		Therefore, we build the FRL framework based on a PG-based algorithm.
		
		PG-based methods directly optimize the policy of agents to maximize the long-term reward, i.e., each agent $k$ aims to maximize its accumulative reward  $R^{(k)}(\tau):=\sum_{t=1}^T R_t^{(k)}$  over trajectory $\tau$, namely 
	   \begin{equation}	       
   \max y_k(\boldsymbol{\Phi}) := \mathbb{E}_{\tau}(R^{(k)}(\tau)),
	   \end{equation}	 
		 where $\boldsymbol{\Phi}$ is the joint policy of all agents and trajectory ${\tau := \{(\mathbf{z}_t^{(k)},\mathbf{a}_t^{(k)},R_t^{(k)})|\forall k\in \mathbb{K}, t\in\mathbb{T}\}}$ is collected by implementing the joint policy of all agents and interacting with environment for a time period of $T$.
		
		We use a fully ReLU-based multi-layer neural network as an approximator of the policy. 
		For agent $k$, we denote its trainable parameters as $\boldsymbol{\theta}_k$ (i.e.,  networks' weights and biases to be trained). 
		The corresponding policy is then expressed as $\pi_k(\mathbf{a}^{(k)}_t|\mathbf{z}_t^{(k)};\boldsymbol{\theta}_k)$ (i.e.,   networks' outputs). 
		After interacting with the environment for $T$ time slots, each agent $k$ can calculate its approximate PG as follows \cite{leonardos2022global},
		\begin{equation}\label{PG}
			\begin{aligned}
				\nabla_{\boldsymbol{\theta}_k} y_k(\boldsymbol{\Phi}) 
				\approx \mathbb{E}_{\tau}\Big(R^{(k)}(\tau) \sum_{t=1}^T\nabla \log \pi_k(\mathbf{a}^{(k)}_t|\mathbf{z}_t^{(k)};\boldsymbol{\theta}_k)\Big).
			\end{aligned}
		\end{equation}		
		With the above PG, the gradient ascent-based optimizers (e.g., RMSprop \cite{hinton2012neural}) can be used to train the networks.

  		\begin{algorithm}
			\caption{RIFRL} 
			\begin{algorithmic}[1] \label{RIFRL}
				\STATE	 \textbf{Input}: Initial global model $\th_{g}^0$, number of training episodes $J$, and aggregation interval $G$.	\\
				
				\FOR{episode index $j = 0, 1, \ldots, J$} 
				\STATE \texttt{--Local learning steps--} 
				\FOR{each agent $k \in {\K}$ in parallel}
				\STATE Update its local model by $\boldsymbol{\theta}_k^{j} = \tilde{\boldsymbol{\theta}}_{g}^{j}$.
				\STATE Sample a trajectory $\tau_{k}^j$ based on current policy $\pi_k(.;\boldsymbol{\theta}_{g}^j)$ and calculate the PG  based on \eqref{PG}.
				\STATE Update $\th_k^{j+1}$ by applying a gradient ascent-based optimizer to train the neural networks.	
				\IF{$j \text{ mod } G ==0 $}
				\STATE Rescale $\th_k^{j+1}$ by \textbf{BRIO} to get $\widetilde{\th}_k^{j+1}$.
				\STATE  Upload  rescaled model $\widetilde{\th}_k^{j+1}$ to the server.
				\ENDIF
				\ENDFOR 
				\STATE \texttt{--Global aggregation--} 
				\IF{$j \text{ mod } G ==0$} 
				\FOR{ the central server}
				\STATE Update  global model $\th_g^{j+1} = \frac{1}{K}\sum_{k=1}^K\widetilde{\th}_{k}^{j+1}$.
				\STATE  Rescale  $\th_g^{j+1}$ by \textbf{BRIO} to get $\widetilde{\th}_g^{j+1}$
				\STATE Broadcast rescaled model $\widetilde{\th}_g^{j+1}$  to all agents.
				\ENDFOR 
				\ENDIF
				\ENDFOR 		
				\STATE \textbf{Output}: Trained global model, $\th_g^{J+1}$.
			\end{algorithmic}
		\end{algorithm}
		
			\subsection{ RIFRL algorithm}
		  The RIFRL algorithm is presented in Algorithm \ref{RIFRL}.
		  In this FRL system, we treat $K$ agents as edge nodes, place a central server to aggregate local models $\{\boldsymbol{\theta}_k:\forall k\in \mathbb{K}\}$ from agents, and broadcast the aggregated global model, $\boldsymbol{\theta}_g$, to all agents.  
		All agents update their local models in each episode and aggregation happens every $G$ episode.
		At Step 5,  local updates for all agents begin with the same models, $\boldsymbol{\theta}_k^{j} = \tilde{\boldsymbol{\theta}}_{g}^{j}$.
		In Step 6, each agent interacts with the environment based on the policy represented by their local models, obtains the experience trajectory of the current episode, and calculates the corresponding PG using \eqref{PG}.
		In Step 7, each agent updates the local model using the obtained PG and the gradient ascent method.
		At Step 16, the central server aggregates the knowledge of all agents.
		Furthermore, we improve the FRL algorithm by adding two backward rescale steps before and after the aggregation step in Steps 9 and 16.
		The BRIO is introduced in the next subsection.
  
  Note that although the FRL algorithm targets a global shared model, only model parameters are shared among all agents and the same model can still output a unique policy for each agent based on different observations of different agents.

		\subsection{Rescale for ReLU activated linear layers}
\begin{figure}[th]
			\centering
			\includegraphics[width=0.40\textwidth]{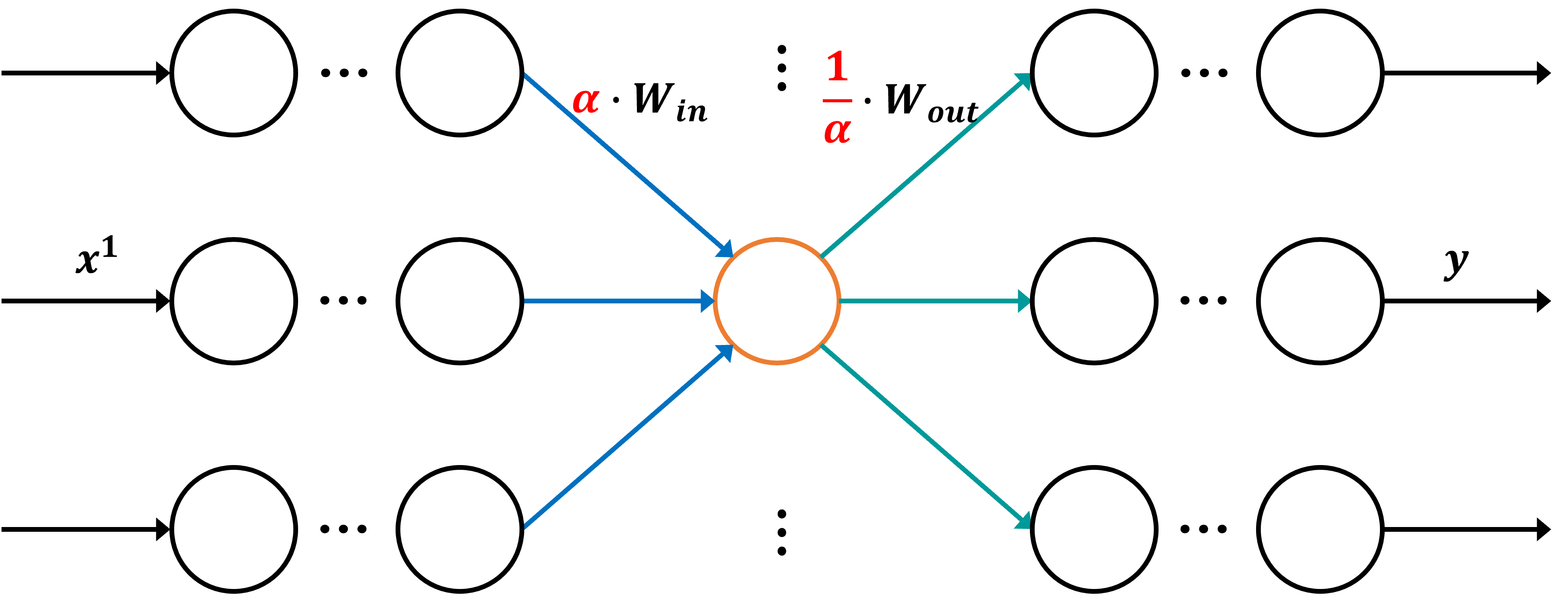}
			\caption{Node-wise invariance.}
			\label{node_wise_invariance}
		\end{figure}

		As shown in Fig. \ref{node_wise_invariance}, node-wise invariance is based on the fact that for a ReLU-activated neuron, if we multiply the input weights of this node by $\alpha>0$ and multiply the output weights of this node by $\frac{1}{\alpha}$, then the input-output map of this node does not change.	Now we integrate this idea 
		in a fully connected network with $L-1$ ReLU-activated linear layers. 
	
	Let the input, output, weight matrices, and bias of the ReLU activated network be $\mathbf{x}^1$, $\mathbf{y}$, $(\mathbf{W}^1,\ldots,\mathbf{W}^L)$ and $(\mathbf{b}^1,\ldots,\mathbf{b}^L)$, respectively. Then we have the input-output relationships as follows,
			\begin{equation}\label{lth_layer}
			\begin{aligned}	
				\mathbf{x}^{\ell+1}&=ReLU(\mathbf{W}^\ell\mathbf{x}^\ell+\mathbf{b}^\ell),~ \ell=1,\ldots,L-1,\\[.5ex]
		 \mathbf{y} &= \mathbf{W}^L\mathbf{x}^L+\mathbf{b}^L.
			\end{aligned}\end{equation}
			Left multiplying a diagonal matrix $\mathbf{D}^{l+1}$ with positive diagonal entries to both sides of \eqref{lth_layer} yields that 
			\begin{equation}\label{mid_net}
			\begin{aligned}
			\widetilde{\mathbf{x}}^{\ell+1}&:=\mathbf{D}^{\ell+1}\mathbf{x}^{\ell+1}\\[.5ex]
			&=\mathbf{D}^{\ell+1} ReLU(\mathbf{W}^\ell\mathbf{x}^\ell+ \mathbf{b}^\ell)\\[.5ex]
			&=ReLU(\mathbf{D}^{\ell+1}\mathbf{W}^\ell\mathbf{x}^\ell+\mathbf{D}^{\ell+1}\mathbf{b}^\ell)\\[.5ex]
			&= ReLU(\mathbf{D}^{\ell+1}\mathbf{W}^\ell(\mathbf{D}^{\ell})^{-1}\widetilde{\mathbf{x}}^\ell+\mathbf{D}^{\ell+1}\mathbf{b}^\ell),\\[.5ex]
			\mathbf{D}^{L}\mathbf{y} &= \mathbf{D}^{L+1}\mathbf{W}^L(\mathbf{D}^{L})^{-1}\widetilde{\mathbf{x}}^L+\mathbf{D}^{L+1}\mathbf{b}^L,
			\end{aligned}		
			\end{equation}
			where the third equation is due to the homogeneity\footnote{A function $f$ is homogeneous if $f(\alpha x)=\alpha f(x)$ for any $\alpha>0$.} of ReLU and the last two equations are from  $\widetilde{\mathbf{x}}^{\ell}=\mathbf{D}^{\ell}\mathbf{x}^{\ell}$. The above equations indicate that we obtain another network with $\widetilde{\mathbf{x}}^\ell$ being the input vector of the $\ell$-th linear layer.
	 		The input and output of the new network are $\widetilde{\mathbf{x}}^1$ and  $\mathbf{\mathbf{D}}^{L+1}\mathbf{y}$.
		 	To make the two networks equivalent, we set $\mathbf{D}^1$ and $\mathbf{D}^{L+1}$ to be identity matrices. 
	 		Then we get a new equivalent network with the $\ell$-th layer's weight matrix and bias vector being 
	 		\begin{equation}\label{new-W}	\widetilde{\mathbf{W}}^\ell:=\mathbf{D}^{\ell+1}\mathbf{W}^\ell(\mathbf{D}^{\ell})^{-1} ~~\text{and}~~\widetilde{\mathbf{b}}^\ell:=\mathbf{D}^{\ell+1}\mathbf{b}^\ell.
	 		\end{equation}
	 		To choose a proper diagonal matrix $\mathbf{D}^{\ell}$, we adopt the idea to normalize $\widetilde{\mathbf{W}}^\ell$ to have unit columns. To fulfil that, we can calculate the $i$-th diagonal entry of matrix $\mathbf{D}^{\ell}$ in a backward rescale order of $\ell=\{L,\ldots,1\}$ as follows,
	\begin{equation}\label{rescale-W}		
		\mathbf{D}^{\ell}_{ii}
		=\left\{\begin{array}{ll}
	 1,&\ell\in\{1, L+1\},\\[1ex]
	\|(\mathbf{D}^{\ell+1}\mathbf{W}^\ell)_{:i}\|, &\ell\in\{2,3,\ldots,L\},
		\end{array}\right.	
		\end{equation}
		where   $\|(\mathbf{D}^{\ell+1}\mathbf{W}^\ell)_{:i}\|$ denotes the Euclidean norm of the $i$-th column of $\mathbf{D}^{\ell+1}\mathbf{W}^\ell$. One can observe that the above choice of $\mathbf{D}^{\ell}$ normalizes each column of $\mathbf{D}^{\ell+1}\mathbf{W}^\ell$ so that  $\widetilde{\mathbf{W}}^\ell$ has unit columns. We call such an operation the backward rescale-invariant operation (BRIO) and present it in Algorithm \ref{backward_rescaling}.
		\begin{algorithm}
			\caption{{\bf BRIO}}
			\begin{algorithmic}[1] \label{backward_rescaling}
				\STATE	 \textbf{Initialize}: A fully connected neural network with $L-1$ ReLU activated linear layers and a linear output layer with weight matrices $(\mathbf{W}^1,\ldots,\mathbf{W}^L)$ and bias $(\mathbf{b}^1,\ldots,\mathbf{b}^L)$. \\
				 
				\STATE For $\ell=L$ to $1$, calculate  $(\mathbf{D}^{L+1},\ldots,\mathbf{D}^{1})$ by \eqref{rescale-W}.	 	
				\STATE Return rescaled weight matrices $(\widetilde{\mathbf{W}}^1,\ldots,\widetilde{\mathbf{W}}^L)$ and bias $(\widetilde{\mathbf{b}}^1,\ldots,\widetilde{\mathbf{b}}^L)$  by \eqref{new-W}.	
			\end{algorithmic}
		\end{algorithm}

To integrate	BRIO into RIFRL, we let $\th:=(\mathbf{W}^1,\ldots,\mathbf{W}^L,$ $\mathbf{b}^1,\ldots,\mathbf{b}^L)$ and the rescaled model $\widetilde{\th}:=(\widetilde{\mathbf{W}}^1,\ldots,\widetilde{\mathbf{W}}^L,$ $\widetilde{\mathbf{b}}^1,\ldots,\widetilde{\mathbf{b}}^L)$. Thanks to this, for each agent $k$, the components in $\widetilde{\th}_k$ corresponding to $\widetilde{\mathbf{W}}^\ell$ have unit columns so that the main components in $\th_k$ corresponding to $\mathbf{W}^\ell$ for different agents become similar, which is beneficial to the model aggregation and thus improves the learning process.

\begin{remark}	To end this section, we would like to point out that the BRIO can be extended to neural networks with other activation functions if they are homogeneous. This is because equations (\ref{mid_net}) still hold if we replace ReLU with these functions. Typical homogeneous activation functions are piecewise linear functions such as leaky RuLU \cite{maas2013rectifier} and parametric ReLU\cite{he2015delving}. Therefore, the  RIFRL algorithm can be flexibly adjusted based on other neural networks with homogeneous activation functions.
	\end{remark}
 
		\section{Simulation Results}\label{simulation}
		In this section, we demonstrate the performance of our proposed RIFRL algorithm for the resource allocation problem in a V2X system.
		We follow the urban case in Annex A of \cite{3GPP_TR_36.885_V14.0.0}, which simulates 9 blocks in total with an area size of 1299m$\times$750m and with both line-of-sight and non-line-of-sight channels. 
		The corresponding channel parameters are set the same as Table I and Table II in \cite{liang2019spectrum}.
		We consider 8 vehicles and thus 8 V2I links as well as 8 sub-channels. 
		The number of V2V links, $K$, is set to 24 and  V2V package data size $D_s$ is set to 2120 Bytes if not specified.

		The policy neural network for each V2V link consists of three fully connected hidden layers with $500$, $250$, and $120$ neurons, respectively.
		ReLU is employed as the activation function in all layers except for the outermost layer where the softmax function is adopted so that the policy network can output a probability vector for the action candidates. To train the network, we take advantage of the RMSprop optimizer.
		
		We compare three learning methods and a random baseline that randomly selects actions and is used to show the system performance without optimization. The three learning methods are the independent PG algorithm, the FedAvg-based FRL algorithm (short for FRL), and  RIFRL.
		We set the learning rate for  FRL  and  RIFRL as $10^{-3}$ but for independent PG as $10^{-4}$.
		This is because a slightly larger learning rate makes independent PG fail to learn a good policy.
		We set the time duration of each time slot as 1 ms and each training episode consists of 100 time slots.
		Both FRL and RIFRL aggregate agents' models once every 5 training episodes, i.e., $G=5$.

		\begin{figure}[t]
			\centering
			\includegraphics[width=0.5\textwidth]{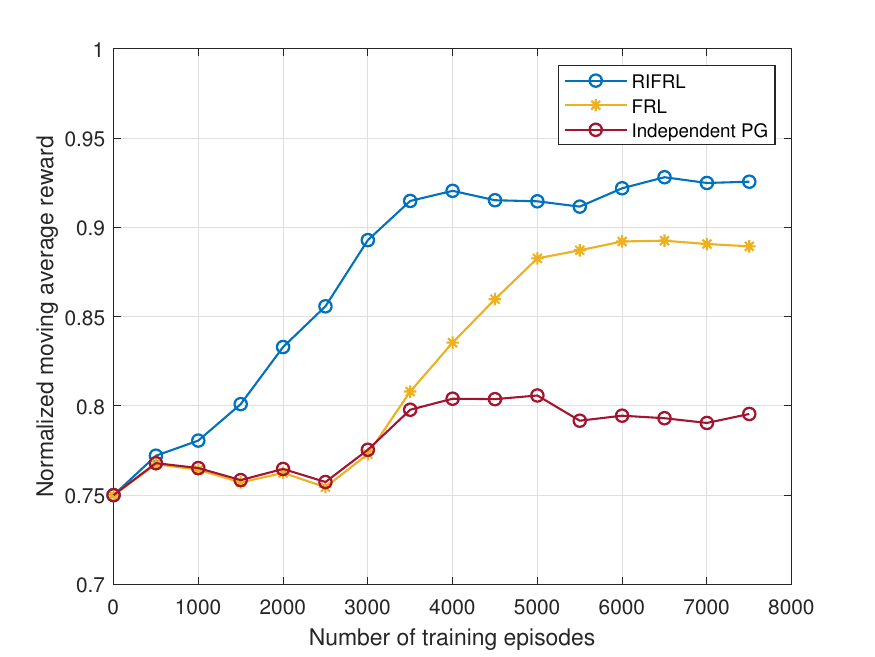}
			\caption{Moving average reward of 3 algorithms.}\vspace{-0mm}
			\label{convergence}
		\end{figure}

		We begin by comparing and analysing the learning processes of three algorithms. Fig. \ref{convergence} illustrates the normalized moving average reward throughout the training phase. In comparison with independent PG,  two FRL algorithms can enhance the learning process by empowering each agent to acquire knowledge beyond its individual observations through interactions with other agents.  This is because the V2V delivery probability is a long-term optimization target. Thus it is challenging for independent PG using a single agent to explore the relation between the long-term reward and the instantaneous local observations.

Among the three algorithms,   RIFRL improves both the learning speed and convergent performance remarkably. These improvements stem from the BRIO, which adjusts the local models of agents to make them more similar to each other, thereby shrinking the solution space of each agent's optimization problem and enhancing the aggregation performance.
We highlight that RIFRL is able to achieve comparable performance to  FRL with around 2000 fewer training episodes.

\begin{figure}[t]
			\centering
			\includegraphics[width=0.5\textwidth]{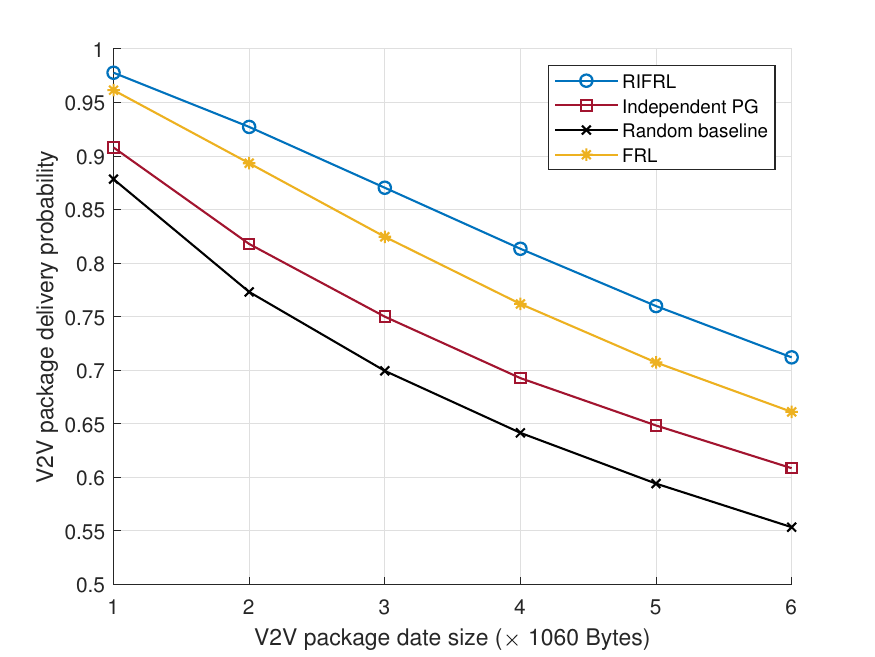}
			\caption{V2V delivery probability of 4 algorithms.}
			\label{datasize}
		\end{figure}

		Next, we evaluate the trained models of all algorithms by employing them in another testing environment. Fig. \ref{datasize} illustrates the performance of different algorithms in terms of V2V packet delivery probability when varying the size of V2V data package. It is noticeable that the probability decreases as the data size increases for all algorithms. It is clear that RIFRL outperforms the other algorithms while both FRL algorithms display superior performance compared to independent PG. 
		
\begin{figure}[t]
			\centering
			\includegraphics[width=0.5\textwidth]{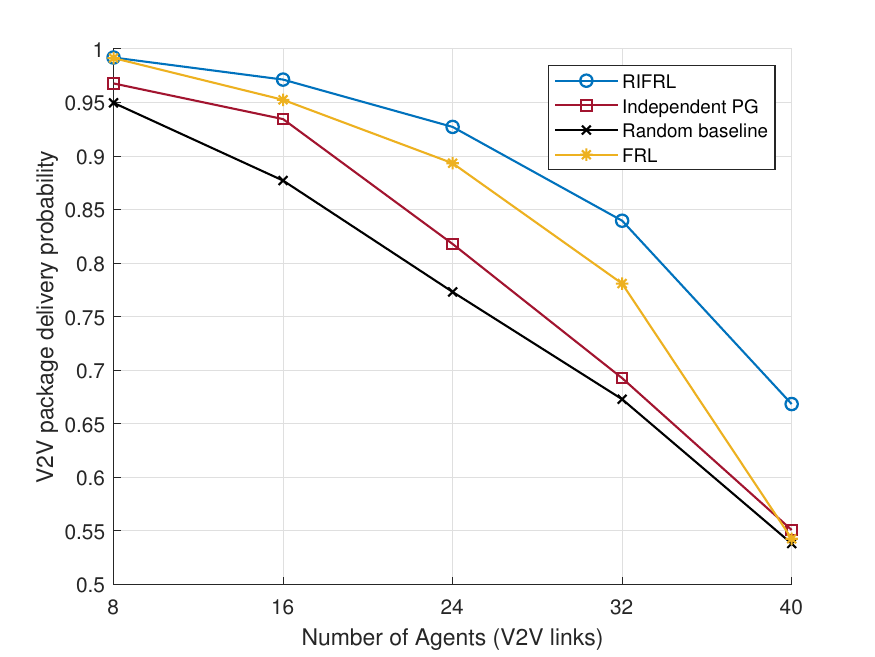}
			\caption{V2V delivery  probability of 4 algorithms.}
			\label{N_agents}
		\end{figure}

		Finally, to assess the adaptation ability of the algorithms under varying environmental configurations, we modify the number of V2V links (i.e., the number of agents), which impacts the V2V link density per resource block and consequently results in diverse environmental statistics. As illustrated in Fig. \ref{N_agents}, it becomes evident that as the number of V2V links increases, the V2V package delivery probability decreases, indicating poorer performance. This degradation is attributed to the increased competition among V2V links sharing the same sub-channel. Once again, RIFRL exhibits the highest probability, highlighting its desirable effectiveness in adapting to different environmental configurations.

		\section{Conclusion}\label{conclusion}

The performance enhancement for resource allocation problems in V2X networks arises from two aspects: the integration of FL and RL and the exploitation of node-wise invariance of ReLU-activated neural networks. This article enables us to design a useful procedure BRIO. We believe that the developed architecture and the BRIO hold the potential to address many other applications, which deserve further investigation.

		\bibliographystyle{IEEEtran}
		\bibliography{bib}

\begin{thebibliography}{10}
\providecommand{\url}[1]{#1}
\csname url@samestyle\endcsname
\providecommand{\newblock}{\relax}
\providecommand{\bibinfo}[2]{#2}
\providecommand{\BIBentrySTDinterwordspacing}{\spaceskip=0pt\relax}
\providecommand{\BIBentryALTinterwordstretchfactor}{4}
\providecommand{\BIBentryALTinterwordspacing}{\spaceskip=\fontdimen2\font plus
\BIBentryALTinterwordstretchfactor\fontdimen3\font minus
  \fontdimen4\font\relax}
\providecommand{\BIBforeignlanguage}[2]{{%
\expandafter\ifx\csname l@#1\endcsname\relax
\typeout{** WARNING: IEEEtran.bst: No hyphenation pattern has been}%
\typeout{** loaded for the language `#1'. Using the pattern for}%
\typeout{** the default language instead.}%
\else
\language=\csname l@#1\endcsname
\fi
#2}}
\providecommand{\BIBdecl}{\relax}
\BIBdecl

\bibitem{ye2019deep}
H.~Ye, G.~Y. Li, and B.-H.~F. Juang, ``Deep reinforcement learning based
  resource allocation for v2v communications,'' \emph{IEEE Trans. Veh.
  Technol.}, vol.~68, no.~4, pp. 3163--3173, 2019.

\bibitem{liang2019spectrum}
L.~Liang, H.~Ye, and G.~Y. Li, ``Spectrum sharing in vehicular networks based
  on multi-agent reinforcement learning,'' \emph{IEEE J. Sel. Areas Commun.},
  vol.~37, no.~10, pp. 2282--2292, 2019.

\bibitem{he2020resource}
Z.~He, L.~Wang, H.~Ye, G.~Y. Li, and B.-H.~F. Juang, ``Resource allocation
  based on graph neural networks in vehicular communications,'' in \emph{2020
  IEEE Glob. Commun. Conf.}, 2020, pp. 1--5.

\bibitem{nguyen2019distributed}
K.~K. Nguyen, T.~Q. Duong, N.~A. Vien, N.-A. Le-Khac, and L.~D. Nguyen,
  ``Distributed deep deterministic policy gradient for power allocation control
  in d2d-based v2v communications,'' \emph{IEEE Access}, vol.~7, pp.
  164\,533--164\,543, 2019.

\bibitem{qi2021federated}
J.~Qi, Q.~Zhou, L.~Lei, and K.~Zheng, ``Federated reinforcement learning:
  Techniques, applications, and open challenges,'' \emph{arXiv preprint
  arXiv:2108.11887}, 2021.

\bibitem{lu2021dynamic}
Z.~Lu, C.~Zhong, and M.~C. Gursoy, ``Dynamic channel access and power control
  in wireless interference networks via multi-agent deep reinforcement
  learning,'' \emph{IEEE Trans. Veh. Technol.}, vol.~71, no.~2, pp. 1588--1601,
  2021.

\bibitem{li2022federated}
X.~Li, L.~Lu, W.~Ni, A.~Jamalipour, D.~Zhang, and H.~Du, ``Federated
  multi-agent deep reinforcement learning for resource allocation of
  vehicle-to-vehicle communications,'' \emph{IEEE Trans. Veh. Technol.}, 2022.

\bibitem{mcmahan2017communication}
B.~McMahan, E.~Moore, D.~Ramage, S.~Hampson, and B.~A. y~Arcas,
  ``Communication-efficient learning of deep networks from decentralized
  data,'' in \emph{Artif. Intell. Stat.}\hskip 1em plus 0.5em minus 0.4em\relax
  PMLR, 2017, pp. 1273--1282.

\bibitem{li2020federated}
T.~Li, A.~K. Sahu, M.~Zaheer, M.~Sanjabi, A.~Talwalkar, and V.~Smith,
  ``Federated optimization in heterogeneous networks,'' \emph{Proc. Mach.
  Learn. Syst.}, vol.~2, pp. 429--450, 2020.

\bibitem{zhou2023fedgia}
S.~Zhou and G.~Y. Li, ``{FedGiA}: {A}n efficient hybrid algorithm for federated
  learning,'' \emph{IEEE Trans. Signal Process.}, vol.~71, pp. 1493--1508,
  2023.

\bibitem{neyshabur2015path}
B.~Neyshabur, R.~R. Salakhutdinov, and N.~Srebro, ``Path-sgd: Path-normalized
  optimization in deep neural networks,'' \emph{Advances in neural information
  processing systems}, vol.~28, 2015.

\bibitem{leonardos2022global}
S.~Leonardos, W.~Overman, I.~Panageas, and G.~Piliouras, ``Global convergence
  of multi-agent policy gradient in markov potential games,'' in \emph{Int.
  Conf. Learn. Rep.}, 2022.

\bibitem{hinton2012neural}
G.~Hinton, N.~Srivastava, and K.~Swersky, ``Neural networks for machine
  learning lecture 6a overview of mini-batch gradient descent,'' \emph{Cited
  on}, vol.~14, no.~8, p.~2, 2012.

\bibitem{maas2013rectifier}
A.~L. Maas, A.~Y. Hannun, A.~Y. Ng \emph{et~al.}, ``Rectifier nonlinearities
  improve neural network acoustic models,'' in \emph{Proc. ICML}, vol.~30,
  no.~1.\hskip 1em plus 0.5em minus 0.4em\relax Atlanta, GA, 2013, p.~3.

\bibitem{he2015delving}
K.~He, X.~Zhang, S.~Ren, and J.~Sun, ``Delving deep into rectifiers: Surpassing
  human-level performance on imagenet classification,'' in \emph{Proc. IEEE
  Int. Conf. Comput. Vis.}, 2015, pp. 1026--1034.

\bibitem{3GPP_TR_36.885_V14.0.0}
{3GPP Technical Specification Group Radio Access Network}, ``Study on
  {LTE}-based {V2X} services; ({Release 14}),'' 3GPP, Technical Report TR
  36.885, June 2016.

\end{thebibliography}

\newpage

%
%
%
%
%
%
%
%

\end{document}